\documentclass[aps,prl,preprint,superscriptaddress]{revtex4}

\begin{document}


\title{An example of Kaluza-Klein-like theory with boundary conditions, which lead to
massless and mass protected spinors chirally coupled to gauge fields}
\author{N. Manko\v c Bor\v stnik}
\affiliation{Department of Physics, University of
Ljubljana, Jadranska 19, SI-1111 Ljubljana}
\altaffiliation{Primorska Institute for Natural Sciences and Technology,
C. Mare\v zganskega upora 2, SI-6000 Koper}
\author{H. B. Nielsen}
\affiliation{Department of Physics, Niels Bohr Institute,
Blegdamsvej 17, Copenhagen, DK-2100}


\author{}
\affiliation{}


\date{\today}

\begin{abstract}
The genuine Kaluza-Klein-like theories (with no fields in addition to gravity) have difficulties 
with the existence of massless spinors after the compactification
of some of dimensions of space\cite{witten}. We assume a $M^{(1+3)} \times$ a flat finite disk  
in $(1+5)$-dimensional space, with the boundary allowing spinors of only one handedness. 
Massless spinors then chirally couple to the corresponding background gauge gravitational field, which solves
equations of motion for a free field, linear in the Riemann curvature.

\end{abstract}


\maketitle

{\it Introduction:}
{\it Genuine Kaluza-Klein-like theories}, assuming nothing but  a gravitational field 
in $d$-dimensional space (no additional gauge or scalar fields), 
which after the spontaneous compactification of a $(d-4)$-dimensional part of space manifest in 
four dimensions as all the known gauge fields including gravity, have difficulties\cite{witten} with 
masslessness of fermionic fields at low energies. It looks  
namely very difficult to avoid after the compactification of a part of space  the appearance of  
representations of both handedness in this part of space and consequently also in the 
(1+3)-dimensional space. Accordingly, the gauge fields can hardly couple chirally in the (1+3) - dimensional 
space. 

In more popular versions, in which one only uses the idea of extra dimensions but does not use
gravity fields themselves to make gauge fields, by just having gauge fields from outset, the break of the parity 
symmetry in the compactified part of space is achieved, for instance, by (an outset of) magnetic fields\cite{salamsezgin}. 
Since gravity does not violate parity, also typically not in extra dimensions alone, it looks accordingly 
impossible to make the genuine Kaluza-Klein gauge particles coupled chirally\cite{holgercolinbook,witten}.
The most popular string theories, on the other side, have such an abundance of ``fundamental`` (or rather 
separate string states) gauge fields, that there is (absolutely) no need for the genuine Kaluza-Klein ones.

In an approach by one of us\cite{pikanorma,norma} 
it has long been the wish to obtain the gauge fields from only gravity, so that ''everything'' would become gravity.
This approach has taken the inspiration from looking for unifying all the internal degrees of freedom, that is 
the spin and all the charges, into only the spin. This approach is also a kind of the genuine Kaluza-Klein theory, suffering
the same problems, with the problem of getting chiral fermions included, unless we can solve them.

There are several attempts in the literature, which use boundary conditions to select massless fields of
a particular\cite{Horawa,Kawamura,Hebecker,Buchmueller}. Boundary conditions are chosen by choosing 
discrete orbifold symmetries.

In this letter we study  a toy model with a Weyl spinor, which caries in $d(=1+5)$ -dimensional 
space with the symmetry $M^{(1+3)}\times $ a flat finite disk only the spin as the internal degree of freedom. 
On the boundary of a finite disk  only  spinors  of one handedness are
allowed. The only back ground field is the gravitational gauge field with vielbeins and spin connections, which
manifest the rotational symmetry on the flat disk. 
We demonstrate that there exist  spinors, which manifest in $M^{(1+3)}$ masslessness (have no partners of opposite
handedness) and are chirally coupled by the Kaluza-Klein charge to the corresponding Kaluza-Klein field. 
The current through the wall is for all the  spinors (the solutions of the Weyl equation
in $d(=1+5)$-dimensional space, which manifest in $d(=1+3)$-dimensional space as massless or massive spinors) 
equal to zero. The Kaluza-Klein charge of all the spinors is proportional to the total angular momentum
on the disk. 

The assumed background field with the spin connections and vielbeins in $d=(1+5)$  (flat on the disk and     
preserving the rotational symmetry on the disk up to gauge transformations) 
fulfills the equations of motion, which follow from the
action in $d=(1+5)$, linear in the Riemann curvature. The action manifests the Kaluza-Klein $U(1)$ gauge field term.

{\it Weyl spinors in gravitational fields with spin connections and vielbeins:} 
We let\cite{hnhep03}\footnote{Latin indices  
$a,b,..,m,n,..,s,t,..$ denote a tangent space (a flat index),
while Greek indices $\alpha, \beta,..,\mu, \nu,.. \sigma,\tau ..$ denote an Einstein 
index (a curved index). Letters  from the beginning of both the alphabets
indicate a general index ($a,b,c,..$   and $\alpha, \beta, \gamma,.. $ ), 
from the middle of both the alphabets   
the observed dimensions $0,1,2,3$ ($m,n,..$ and $\mu,\nu,..$), indices from the bottom of the alphabets
indicate the compactified dimensions ($s,t,..$ and $\sigma,\tau,..$). We assume the signature $\eta^{ab} =
diag\{1,-1,-1,\cdots,-1\}$.
} a spinor interact with a gravitational field through vielbeins
$f^{\alpha}{}_{a}$ (inverted vielbeins to 
$e^{a}{}_{\alpha}$ with the properties $e^a{}_{\alpha} f^{\alpha}{}_b = \delta^a{}_b,\; 
e^a{}_{\alpha} f^{\beta}{}_a = \delta^{\beta}_{\alpha} $ ) and   
spin connections, namely 
$\omega_{ab\alpha}$, which is the gauge field of $S^{ab}= \frac{i}{4}(\gamma^a \gamma^b - \gamma^b \gamma^a)$.
We choose the basic states in the  space of spin degrees of freedom to be 
eigen states of the Cartan sub algebra of the operators: $S^{03}, S^{12}, S^{56}$.

The covariant momentum of a spinor is taken to be
\begin{eqnarray}
p_{0 a} &=& f^{\alpha}{}_{a}p_{0 \alpha}, \quad p_{0 \alpha} \psi = p_{ \alpha} - \frac{1}{2} S^{cd} 
\omega_{cd \alpha}, 
\label{covp}
\end{eqnarray}
when applied to a spinor function $\psi$. 

A kind of a total covariant derivative of $e^a{}_{\alpha}$ (a vector with both-Einstein and 
flat index) is taken to be 
\begin{eqnarray}
p_{0 \alpha} e^{a}{}_{\beta} = i e^a{}_{\beta ; \alpha} = i (e^a{}_{\beta , \alpha} +
\omega^a{}_{d \alpha} e^d{}_{\beta} - 
\Gamma^{\gamma}{}_{\beta \alpha} e^a{}_{\gamma}).
\label{covderviel}
\end{eqnarray}
We require that this derivative of a vielbein is zero: $e^a{}_{\beta ; \alpha} =0.$

The corresponding Lagrange density ${\cal L}$  for   a Weyl has the form
${\cal L} = E \frac{1}{2} [(\psi^{\dagger}\gamma^0 \gamma^a p_{0a} \psi) + (\psi^{\dagger} \gamma^0\gamma^a p_{0 a}
\psi)^{\dagger}]$ and leads to
\begin{eqnarray}
{\cal L} &=& E\psi^{\dagger}\gamma^0 \gamma^a   ( p_{a} - \frac{1}{2} S^{cd}  \Omega_{cda})\psi,
\label{weylL}
\end{eqnarray}
with $ E = \det(e^a{}_{\alpha}) $, 
$ \Omega_{cda} =\frac{1}{2}( \omega_{cda} + (-)^{cda}  \omega^*{}_{cda})$, and with
$(-)^{cda}$, which is $-1$, if two indices are equal, and is $1$ otherwise (if all three indices are different). 
(In $d=2$ case
$\Omega_{abc}$ is always pure imaginary.)

The Lagrange density (\ref{weylL}) leads to the Weyl equation
\begin{eqnarray}
\gamma^0 \gamma^a  P_{0a}\psi =0, \quad P_{0a}=   (f^{\alpha}{}_{a} p_{\alpha} -\frac{1}{2} S^{cd} \Omega_{cda}).
\label{Weylgen}
\end{eqnarray}
Taking now into account that $\gamma^a \gamma^b = \eta^{ab}- 2i S^{ab}$,  $\{\gamma^a, S^{bc}\}_- = i
(\eta^{ab} \gamma^c - \eta^{ac} \gamma^b)$,  one easily finds that\footnote{
 $[a  ...b]$ means, that the expression must be anti symmetrized with respect to $a,b$. }
 $ \gamma^a P_{0a} \gamma^b P_{0b} = P_{0a} P_{0}^{a} 
+ \frac{1}{2} S^{ab}  S^{cd}{\cal R}_{abcd} + 
S^{ab} {\cal T}^{c}{}_{ab} P_{0 c}. $
We find  ${\cal R}_{ab cd} = f^{\alpha}{}_{[a} f^{\beta}{}_{b]} (\Omega_{cd \alpha,\beta } + \Omega_{ce\alpha}
\Omega^{e}{}_{d \beta})$ and for the torsion:
${\cal T}^{c}{}_{ab} =  
 f^{\alpha}{}_{[a} (f^{\beta}{}_{b]})_{, \alpha} e^c{}_{\beta} 
+ \Omega_{[a}{}^{c}{}_{b]}$.

The most general vielbein for $d=2$ can be written by an appropriate parameterization as
\begin{eqnarray}
e^s{}_{\sigma} = e^{\varphi/2}
\pmatrix{\cos \phi   & \sin \phi \cr
- \sin \phi & \cos \phi \cr}, 
f^{\sigma}{}_{s} = e^{-\varphi/2} \;
\pmatrix{\cos \phi   & -\sin \phi \cr
\sin \phi & \cos \phi \cr},
\label{f1and1}
\end{eqnarray}
with $s=5,6$ and $\sigma =(5),(6)$ and $g_{\sigma \tau} = e^{\varphi} \eta_{\sigma \tau}, g^{\sigma \tau} = e^{- \varphi}
\eta^{\sigma \tau}$, $\eta_{\sigma \tau} = diag(-1,-1) = \eta^{\sigma \tau}$.
If there is no dilatation then $ E =1$. 
If in the case of $d=2$, the Einstein action for a free gravitational field 
$S= \int \; d^d{} x \; E \; R ,$ with $  
R = f^{\sigma [s} f^{\tau t]} \;\Omega_{st \sigma,\tau} 
$ is varied 
with respect to both, spin connections and vielbeins, the corresponding equations of motion
bring no conditions on any of these two types of fields,
so that any zweibein and any spin connection can be assumed. 
Accordingly,  if we put the expression in Eq.(\ref{covderviel}) equal to zero, can this 
equation {\it for $d=2$ be  understood as the defining equation 
for ${\cal T}^{s}{}_{\sigma \tau}\equiv \Gamma^{s}{}_{\sigma \tau} - \Gamma^{s}{}_{\tau \sigma}$,}
for any spin connection $\Omega_{stu}$ and any zweibein.

We assume that a two dimensional space is  a flat  disk 
\begin{eqnarray}
f^{\sigma}{}_{s} = \delta^{\sigma}{}_{s},\; \omega_{56 s} =0,
\label{disk}
\end{eqnarray}
with the rotational symmetry and with the radius $\rho_0$. We require that spinors must 
obey the boundary condition
\begin{eqnarray}
(1-i n^{(\rho)}{}_{\alpha} n^{(\varphi)}{}_{\beta} 
f^{\alpha}{}_{a}f^{\beta}{}_{b} \gamma^a \gamma^b )\psi|_{\rho= \rho_0}=0,
\label{diskboundary}
\end{eqnarray}
where $\psi$ is the solution of the Weyl equation in $d=1+5$ and 
$n^{(\rho)}=(0,0,0,0,\cos \varphi, \sin \varphi),\; n^{(\varphi)}= (0,0,0,0,-\sin \varphi, \cos \varphi)$ 
are the two unit vectors 
perpendicular and tangential to the boundary (at $\rho_0$), respectively. We shall see  in what follows that 
the boundary 
forces massless spinors - that is spinors, which manifest masslessness in $(1+3)$-dimensional space - 
to be of only right handedness\footnote{If we start with the left handed spinor in $d=1+5$, the
handedness of a massless spinor in $d=1+3$ is chosen to be the left one.}, 
while the current at the boundary is in the perpendicular direction
equal to zero for either massless or massive spinors. 
Spinors manifest masslessness in $d=(1+3)$-dimensional space, if they solve the Weyl equation (\ref{Weylgen})
with $a=5,6$, so that the term $E\psi^{\dagger}\gamma^0 \gamma^s  p_{0s}\psi,\; s=5,6$ (the only term, which would 
manifest as a mass term in $d=1+3$) is equal to zero.
The boundary condition assures the mass protection. 


{\it Weyl spinors on a flat disk:} The Weyl spinor  wave functions $\psi$, manifesting masslessness
 in the $(1+3)$ space, 
must on a flat disk obey the Weyl equations of motion (Eq.(\ref{Weylgen})), while the Weyl wave functions in $d=1+5$, 
manifesting in $d=1+3$  a mass $m$, obey the  Dirac equation
 \begin{eqnarray}
   \gamma^5  e^{2i \phi S^{56} }(p_{0(5)}  + 
   2i S^{56} p_{0 (6)})\psi = -m \psi.
 \label{weyld=2m}
 \end{eqnarray}
For our flat disk $\phi =0$ and $\omega_{st \sigma} =0.$ The Dirac equation in $d=2$ goes to the Weyl equation,
if we assume $m=0$. The Weyl wave function in $d=1+5$, may (in our case for a (flat disk)  $\times M^{1+3}$)
be written as a product of a function, which depends only on the 
coordinates $x^{\sigma}, \sigma = (5),(6),$
and the part, which we denote by $(+)$ and $[-]$ and includes the dependence on all the internal degrees of freedom
(in (1+5)) as well as on the coordinates  $x^{\mu}, \mu=(0),(1),(2),(3).$ It follows that
$(+)$ and $[-]$ are the two types of 
the spinor states, corresponding to the eigen values of $S^{56}$
equal to $+1/2$ and $-1/2$, respectively (what ever the spinor and coordinate content in $d=1+3$ might be). 
Accordingly $(+)$ represents the right  and $[-]$ the left handed spinor state in $d=2$ ($\Gamma^{(2)} = 2 S^{56}$), while 
in $d=(1+3)$ they represent the left and the right  handed spinors, respectively (since we started in $d=(1+5)$ with the 
left handed Weyl spinor). 

We introduce the polar coordinates in $d=2$: $x^{(5)}=\rho \cos \varphi, \; x^{(6)} = \rho \sin \varphi$ and
require that states are also eigen states of the total angular momentum $M^{56} = -i \frac{\partial}{\partial \varphi}
+ S^{56}$. We then write eigen states of the Dirac equation (\ref{weyld=2m}) as
\begin{eqnarray}
\psi^{m}{}_{+(n+1/2)}(\rho, \varphi) &=& \alpha^{+}{}_{n}(\rho) e^{i n \varphi}(+) + \beta^{+}{}_{n+1}(\rho) e^{i (n+1) 
\varphi} [-],\nonumber\\
\psi^{m}{}_{-(n+1/2)}(\rho, \varphi) &=& \alpha^{-}{}_{n+1}(\rho) e^{-i (n+1) \varphi}(+) + \beta^{-}{}_{n}(\rho) e^{-i n 
\varphi} [-].  
 \label{solutionweylm}
 \end{eqnarray}
Index $m$ denotes that spinors carry a mass $m$. The functions $\alpha^{\pm}{}_{k}, \; \beta{\pm}{}_{k},\; 
k=n, n+1, $ $n=0,1,2,3...,$ must solve Eq.(\ref{weyld=2m}) (for either 
the massive or the  massless case), which in the polar coordinates read
 \begin{eqnarray}
i (\frac{\partial}{\partial \rho} - \frac{n}{\rho}) \alpha^{+}{}_{n}(\rho) &=& m \;\beta^{+}{}_{n+1} (\rho),\nonumber\\ 
i (\frac{\partial}{\partial \rho} + \frac{n+1}{\rho}) \beta^{+}{}_{n+1} (\rho) &=& m \;\alpha^{+}{}_{n}(\rho).
 \label{weyld=2mpolar}
 \end{eqnarray}
One immediately finds that for $m=0$ the functions  $\alpha^{+}{}_{n}(\rho) = \rho^n$, 
$\beta^{+}{}_{n+1} = 0$ and $\alpha^{-}{}_{n+1}=0,$ $\beta^{-}{}_{n} =\rho^n$, $n=0,1,2...,$ solve 
Eq.(\ref{weyld=2mpolar}), 
so that the right handed solutions and the left handed solutions in $d=2$ are as follows
\begin{eqnarray}
\psi^{m=0}{}_{+(n+1/2)}(\rho, \varphi) = A_{+(n+\frac{1}{2})} \rho^{n} e^{i n \varphi}(+), \nonumber\\
\psi^{m=0}{}_{-(n+1/2)}(\rho, \varphi) = A_{-(n+\frac{1}{2})} \rho^{n} e^{-i n \varphi} [-], 
 \label{solutionweylm=0}
 \end{eqnarray}
 with $A_{\pm(n+\frac{1}{2})}$  constants.
In the massive case, for the two types of functions $\alpha^{\pm}{}_{m}$ and $\beta^{\pm}{}_{m}$  
the Bessel functions can be taken as follows:
$\alpha^{+}{}_{n}(\rho)=\beta^{-}{}_{n}(\rho) = J_n $ and $\beta^{+}{}_{n+1} (\rho)= \alpha^{-}_{n+1}(\rho) = J_{n+1}$.
The zeros of the Bessel functions are at $m \rho_0 = 2,4..,3.8..,5.1..,etc, $ corresponding to  $J_0,J_1,J_2,...$, 
respectively.


{\it Boundary conditions:} According to the boundary condition of Eq.(\ref{diskboundary}), which in our case 
indeed requires that
\begin{eqnarray}
0= (1-\Gamma^{(2)}) \psi^{m}{}_{\pm(n +1/2)}|_{\rho=\rho_{0}}, 
\label{diskboundary1}
\end{eqnarray}
only the masses, for which $\beta^{\pm}{}_{k} (\rho_0)=0$, are allowed, since the term with 
$(+)$ is at $\rho_0$ multiplied by zero, while the term with $[-]$ is multiplied by (1+1).
In the massless case, the boundary condition requires that $A_{-(n+\frac{1}{2})} =0$, so that only right handed
spinors with the spin part $(+)$ survive. 
There are accordingly infinite number of massive and  of massless solutions. To different solutions different
total angular moments correspond and in the massive case also different masses.

We easily see that a current through the wall 
\begin{eqnarray}
n^{(\rho)}{}_{\alpha} j^{\alpha}|_{\rho= \rho_0} = \psi^+ \gamma^0 \gamma^a f^{\alpha}{}_{a} n^{(\rho)}{}_{\alpha}
\psi|_{\rho=\rho_0}, 
\label{diskcurrentthroughwall}
\end{eqnarray}
is in all the cases (massless and massive) equal to zero. In the massive case, the current  is proportional to the terms
$\alpha^{\pm}{}_{k} (\rho_0) \beta^{\pm}{}_{k \pm 1}(\rho_0)$, which are zero, since always
either $\alpha^{\pm}{}_{k} (\rho_0)$ or $\beta^{\pm}{}_{k \pm 1}(\rho_0)$ is zero on the wall.
In the massless case $\beta^{+}{}_{m}$ is zero all over.


{\it Spinors coupled to gauge fields in $M^{(1+3)} \times M^{(2)}$, with $M^{(2)}$ a flat finite disk:} 
To study how do spinors couple to the Kaluza-Klein gauge fields in the case of $M^{(1+5)}$, ''broken'' to 
$M^{(1+3)} \times $ a flat disk with $\rho_0$ and the boundary condition, which allows only right handed spinors
at $\rho_0$,
we first look for (background) gauge gravitational fields, which preserve the rotational symmetry on the disk.
To find such fields, we study the coordinate transformations  of the type $x^{' \mu}= x^{\mu}, x^{' \sigma}= x^{\sigma}
+ \zeta^{\sigma}(x^{\tau}) \theta (x^{\mu})$, with\footnote{$\varepsilon^{(5)}{}_{(6)} =-1=
- \varepsilon^{(6)}{}_{(5)}$, while the rest of terms are zero.} $\zeta^{\sigma}(x^{\tau})=
\zeta^{\sigma}{}_{0} + \varepsilon^{\sigma}{}_{\tau}x^{\tau}$. We start with $f^{\alpha}{}_{a} = \delta^{\alpha}{}_{a},
\omega_{ab \alpha} = 0.$  Requiring globaly that $\delta_{0} f^s{}_{\sigma}= 0$ ($\delta_0 f^{\sigma}{}_{a}=
\omega_{a}{}^{b}\; f^{\sigma}{}_{b} + \zeta^{\sigma}{}_{,\tau}\; \theta(x^{\mu}) \;f^{\tau}{}_{a} - 
\theta(x^{\mu}) \;\zeta^{\tau} \;f^{\sigma}{}_{a ,\tau}$) 
and $\delta_{0} \omega_{st \sigma}=0$  ($\delta_{0} \omega_{st \mu}  = \omega_{s}{}^{a}\; \omega_{a t \mu} + 
\omega_{t}{}^{b}\; \omega_{s b \mu}
- \zeta^{\tau}{}_{,\mu}\; \theta(x^{\nu}) \; \omega_{s t \tau} - \theta(x^{\nu})\; \zeta^{\tau} \;\omega_{st \mu, \tau}
-\omega_{st ,\mu}$) we end up with  $\omega_{st ,\sigma}=0, \zeta^{\sigma}{}_{0}=0$ and,  
by replacing $\theta_{,\mu}$ with $A_{\mu}$ (which is the gauge $U(1)$ field  whose gauge transformation leads
to $ A_{\mu} + \theta_{,\mu}$), with  
$\delta_{0} f^{\sigma}{}_{m} = A_{\mu} \delta ^{\mu}{}_{m}
\varepsilon^{\sigma}{}_{\tau} x^{\tau}$, $\delta_{0} \omega_{st \mu} = - \varepsilon_{st} A_{\mu}$. 
Accordingly the following background vielbein field  
\begin{eqnarray}
e^a{}_{\alpha} = 
\pmatrix{\delta^{m}{}_{\mu}  & e^{m}{}_{\sigma}=0 \cr
 e^{s}{}_{\mu} & e^s{}_{\sigma} \cr},
f^{\alpha}{}_{a} =
\pmatrix{\delta^{\mu}{}_{m}  & f^{\sigma}{}_{m} \cr
0= f^{\mu}{}_{s} & f^{\sigma}{}_{s} \cr},
\label{f6}
\end{eqnarray}
with $f^{\sigma}{}_{m} = A_{\mu} \delta ^{\mu}{}_{m}
\varepsilon^{\sigma}{}_{\tau} x^{\tau}$
and  the spin connection field 
\begin{eqnarray}
\omega_{st \mu} = - \varepsilon_{st}  A_{\mu},\quad \omega_{sm \mu} = -\frac{1}{2} F_{\mu \nu} \delta^{\nu}{}_{m}
\varepsilon_{s \sigma} x^{\sigma}
\label{omega6}
\end{eqnarray}
are assumed. The term $\omega_{sm\mu} = -\frac{1}{2} F_{\mu \nu} \delta^{\nu}{}_{m}
\varepsilon_{s \sigma} x^{\sigma} = - \omega_{ms\mu}$, which is  proportional to $F^{\mu \nu} = \partial^{\mu} A^{\nu}
- \partial^{\nu} A^{\mu}$, can not be derived in the above way, since $A^{\mu}$, if pure gauge, contributes zero to it.
But this term and all the others are the solution of the equations of motion, which follow from the action, linear in 
the Riemann curvature, as we shall see bellow.
 The $U(1)$ gauge field $A_{\mu}$ depends only on $x^{\mu}$.
All the other components of the spin connection fields are zero, since for simplicity we allow no gravity in
$(1+3)$ dimensional space.

To determine the current, coupled to the Kaluza-Klein gauge fields $A_{\mu}$, we
analyze the spinor action
\begin{eqnarray}
{\cal S} &=& \int \; d^dx E \bar{\psi} \gamma^a P_{0a} \psi = \int \; 
d^dx  \bar{\psi} \gamma^m \delta^{\mu}{}_{m} p_{\mu} \psi + \nonumber\\
&& \int \; d^dx   \bar{\psi} \gamma^m (-)S^{sm} \omega_{sm \mu} \psi  + 
\int \; d^dx  \bar{\psi} \gamma^s \delta^{\sigma}{}_{s} p_{\sigma} \psi +\nonumber\\
&& \int \; d^dx   \bar{\psi} \gamma^m  \delta^{\mu}{}_{m}A_{\mu} (\varepsilon^{\sigma}{}_{\tau} 
 p_{\sigma} + S^{56}) \psi.
\label{spinoractioncurrent}
\end{eqnarray}
 $\psi$ are defined in $d=(1+ 5)$ dimensional space (as solutions of the Weyl equation) 
and  solve the Dirac (massive - if $\bar{\psi} \gamma^s \delta^{\sigma}{}_{s} 
p_{\sigma} \psi=-m$) or the Weyl (massless - if $\bar{\psi} \gamma^s \delta^{\sigma}{}_{s} 
p_{\sigma} \psi=0$) equation in $d=2$ (the terms $(+)$ and $[-]$ determine the spin part in $d=1+5$ and
also the dependence on $x^{\mu}$). $E$ is for $f^{\alpha}{}_{a}$ from (\ref{f6}) equal to 1. 
The first term on the right hand side  of Eq.(\ref{spinoractioncurrent}) is the kinetic term
(together with the last  term defines  
the  covariant derivative $p_{0 \mu}$ in $d=1+3$).  
The second term on the right hand side  contributes nothing when integration over 
the disk is performed, since it is proportional to $x^{\sigma}$ ($\omega_{sm \mu} = -\frac{1}{2}
F_{\mu \nu} \delta^{\nu}{}_{m} \varepsilon_{s \sigma} x^{\sigma}$).

We end up with 
\begin{eqnarray}
j^{\mu} = \int \; d^2x \bar{\psi} \gamma^m \delta^{\mu}{}_{m} M^{56}  \psi
\label{currentdisk}
\end{eqnarray}
as  the current in $d=1+3$. {\it The charge in $d=1+3$ is obviously  proportional to the total 
angular momentum  $M^{56} =L^{56} + S^{56}$ on a disk}, for either massless or massive spinors.
One notices, that our toy model allows massless spinors of any angular momentum in $d=2$, which then means that
spinors of charges, proportional to $n+1/2$, for any $n$ are allowed.


{\it Gauge fields on $M^{(1+3)} \times$ a finite disk:}
One can check that  the gauge field in Eqs.(\ref{f6},\ref{omega6}) is in agreement with the 
relation between $\omega_{ab\alpha}$ 
and $f^{\alpha}{}_{a}$, which follow when varrying the action for a free gauge field, if linear in the Riemann curvature 
 ($\int d^dx ER$), with respect to the spin connection field $\omega_{ab\alpha}$:
 $\omega_{ab\alpha} = 
 -\frac{1}{2E}\biggl\{
   e_{e\alpha}e_{b\gamma}\,\partial_\beta(Ef^{\gamma[e}f^\beta{}_{a]} )
   + e_{e\alpha}e_{a\gamma}\,\partial_\beta(Ef^{\gamma}{}_{[b}f^{\beta e]})
  {} - e_{e\alpha}e^e{}_\gamma\,
     \partial_\beta\bigl(Ef^\gamma{}_{[a}f^\beta{}_{b]} \bigr)
   \biggr\} 
  - \frac{1}{d-2}  
   \biggl\{ e_{a\alpha} 
            \frac{1}{E} e^d{}_\gamma \partial_\beta
             \left(Ef^\gamma{}_{[d}f^\beta{}_{b]}\right)
               {} - e_{b\alpha} 
            \frac{1}{E} e^d{}_\gamma \partial_\beta
             \left(Ef^\gamma{}_{[d}f^\beta{}_{a]}\right)
               \biggr\}.$
For the particular solution of a general $(1+5)$ case, when $f^{\mu}{}_{m} = \delta^{\mu}{}_{m}$ and
$\omega_{mn \mu} =0$, which concerns the case with no gravity in $(1+3)$ space, the Riemann tensor  simplifies
to $R= -\frac{1}{2} F^{\mu \nu} F_{\mu \nu} x^{\sigma} x_{\sigma}$, which after the integration over  
$\rho \;d\rho d \varphi$ leads to the known action for the gauge $U(1)$ field.
One can check that the boundary term contributes zero.


{\it Conclusions:}
We start with one Weyl spinor of only one handedness in a space $M^{1+5}$,  and assume that
the space factorizes into $M^{(1+3)} \times$ a flat finite disk with the radius $\rho_0$ and 
with the boundary, which 
allows only spinors of a particular handedness: $(1-i n^{(\rho)}{}_{\alpha} n^{(\varphi)}{}_{\beta} 
f^{\alpha}{}_{a}f^{(\beta)}{}_{b} \gamma^a \gamma^b )\psi|_{\rho= \rho_0}=0$. 
The spinor, whose  only internal degree of freedom  
is the spin,  interacts  with the gauge gravitational field represented by  
spin connections ($\omega_{ab \alpha}$) and vielbeins ($f^{\alpha}{}_{a}$). 
The disk (manifesting the rotational symmetry) is flat ($f^{\sigma}{}_{s}=\delta^{\sigma}{}_{s}$, $\omega_{st\sigma}=0$). 
We look for massless  spinors  in $(1+3)$ ''physical'' space, which are mass protected and chirally coupled
to a Kaluza-Klein gauge field through a quantized (proportional to an integer) Kaluza-Klein charge. 

To be massless in $(1+3)$ space, spinors must  obey the Weyl equation on a disk: $\gamma^0\gamma^s 
f^{\sigma}{}_{s} p_{\sigma}\psi=0, s=\{5,6\}, \sigma =\{(5),(6) \}$. 
The boundary condition on the disk makes the current of (massless and massive) spinors in the perpendicular 
direction to be zero and guarantees that  massless spinors are mass protected.

The background gauge field, chosen to obey isometry relations and 
respecting accordingly the rotational symmetry on the disk, fullfils the general equations of motion, which follow 
in $(1+5)$ from the action, linear in the Riemann curvature. The effective Lagrangean in $d=(1+3)$ is for the flat space
the ordinary Largangean for the $U(1)$ field. The current (for massless or massive spinors) is in the
$(1+3)$-dimensional space proportional to the total angular momentum on the disk $(M^{56})$, which 
is accordingly determining the charge of spinors (proportional to $n+1/2, n=0,1,2,..$). 
 Consequently massless spinors are mass protected and chirally coupled to the Kaluza-Klein gauge fields.

The ''real'' case, with  a spin in $d-$dimensional space, which would manifest in (1+3)-dimensional
''physical'' space  the spin and all the known charges,
needs, of course, much more than two additional dimensions.  All the relations are then much more complex. 
But some properties will very likely repeat, like: 
Spinors with only a spin in $(1+ (d-1))$-dimensional
space will manifest in $(1+ (q-1))$-dimensional space the masslessness, together with the mass 
protection and the charge, 
if a kind of boundary conditions
in a compactified $(d-q)-$ dimensional
space would made possible the existence of spinors of only one handedness. 

It would be worthwhile to find out how our boundary conditions are related (if at all) to the boundary conditions, 
which introduce orbifolds\cite{Horawa,Kawamura,Hebecker,Buchmueller}. 

\section*{Acknowledgments} 
It is our pleasure to thank all the participants of the annual workshops entitled ''What comes beyond 
the Standard model'' (taking place at Bled from 1999 to 2005) for the fruitful discussions. In particular 
we would like to stress discussions of one of us(SNMB) with M. Blagojevi\' c on the topic of the gauge gravity.

\end{document}